\documentclass[structabstract]{aa}  

\usepackage{graphicx}
\usepackage{txfonts}
\usepackage{natbib}
\bibpunct{(}{)}{;}{a}{}{,}

\newcommand{\feh}{\mathrm{[Fe/H]}}
\newcommand{\teff}{T_\mathrm{eff}}

\newcommand{\fei}{Fe~\textsc{i}}
\newcommand{\feii}{Fe~\textsc{ii}}
\newcommand{\ms}{m~s$^{-1}$\ }

\defcitealias{kdwarfs-p1}{Paper~I}
\defcitealias{kdwarfs-p2}{Paper~II}
\defcitealias{kdwarfs-p3}{Paper~III}

\begin{document}

\title{Granulation in K-type Dwarf Stars}
\subtitle{I. Spectroscopic observations}
\titlerunning{Granulation in K-Dwarfs. I.}
\authorrunning{Ram\'{\i}rez et~al.}

\author{I. Ram\'{\i}rez\thanks{Current address: Max Planck Institute for Astrophysics, Postfach 1317, 85741 Garching, Germany} \and
        C. Allende Prieto\thanks{Current address: Mullard Space Science Laboratory, University College London, Holmbury St.~Mary, Dorking, Surrey RH5 6NT, UK} \and
        D. L. Lambert
       }

\institute{McDonald Observatory and Department of Astronomy,
           University of Texas, Austin, TX 78712-0259, USA
          }

\date{Received September 4, 2008; accepted October 23, 2008}

\abstract
{}
{We seek to detect and quantify the effects of surface convection (granulation) on the line spectra of K-dwarfs as a first step towards a rigorous testing of hydrodynamic models for their atmospheres.}
{Very high resolution ($R\simeq160,000-210,000$), high signal-to-noise ratio ($S/N\gtrsim300$) spectra of nine bright K-dwarfs were obtained with the 2dcoud\'e spectrograph on the 2.7\,m Telescope at McDonald Observatory to determine wavelength shifts and asymmetries of \fei\ lines. Spectra of the same stars acquired with the High Resolution Spectrograph ($R\simeq120,000$) on the 9.2\,m Hobby Eberly Telescope were used as radial velocity templates to calibrate the wavelength scale of the 2dcoud\'e spectra.}
{The observed shapes and positions of \fei\ lines reveal asymmetries and wavelength shifts that indicate the presence of granulation. In particular, line bisectors show characteristic C-shapes while line core wavelengths are blueshifted by an amount that increases with decreasing equivalent width ($EW$). On average, \fei\ line bisectors have a span that ranges from nearly 0 for the weakest lines (residual core flux~$\gtrsim0.7$) to about 75~\ms for the strongest lines (residual core flux~$\simeq0.3$) while wavelength shifts range from about $-150$~\ms in the weakest ($EW\simeq10$\,m\AA) lines to 0 in the strongest ($EW\gtrsim100$\,m\AA) features. A more detailed inspection of the bisectors and wavelength shifts reveals star-to-star differences that are likely associated with differences in stellar parameters, projected rotational velocity, and stellar activity. While the former two are well understood and confirmed by our data, the relation to stellar activity, which is based on our finding that the largest departures from the expected behavior are seen in the most active stars, requires further investigation. For the inactive, slow projected rotational velocity stars, we detect, unequivocally, a plateau in the line-shifts at large $EW$ values ($EW\gtrsim100$\,m\AA), a behavior that had been identified before only in the solar spectrum. The detection of this plateau allows us to determine the zero point of the convective blueshifts, which is useful to determine absolute radial velocities. Thus, we are able to measure such velocities with a mean uncertainty of about 60~\ms for four of our sample stars.}
{}


\maketitle

\section{Introduction}

The atmospheric structures of cool stars with convective envelopes are inhomogeneous due to the interplay between the radiation field and the thermodynamic properties of the gas. Near the surface, radiation losses and the strong temperature sensitivity of the continuum opacity induce a very steep temperature gradient, driving convection. The ascending cells of gas (granules) cool down quickly and the gas falls back into the stellar interior through filamentary structures (intergranular lanes). The downflows are filamentary due to mass conservation and they produce the buoyancy work necessary to support the upflows \citep[e.g.,][]{stein98}. This phenomenon, \textit{granulation}, is directly observed in spatially resolved images and spectrograms of the solar disk. For distant stars, and also for the Sun, the signatures of granulation can be detected in their spectra using precise measurements of absorption line profiles \citep[e.g.,][]{gray82,gray05,gray85,gray89,dravins87:line_asymmetries,dravins08,asplund00:iron_shapes,allende02}.

The high degree of correlation between the intensity and velocity fields in inhomogeneous stellar atmospheres produces asymmetric lines (in particular when the stellar disk is unresolved), with characteristic shapes for the \textit{line bisectors} (the midpoints of the horizontal segments across the wings of the lines). In addition, the wavelengths of spatially unresolved lines are blueshifted, owing to the larger contribution to the emergent flux from the brighter, hotter ascending gas. These Doppler shifts are often referred to as \textit{convective blueshifts}. Note, however, that not all lines seen in a stellar spectrum suffer from these effects but only those that are formed in layers that are deep enough into the stellar atmosphere, where the correlation between temperature and velocity fields due to granulation is strong. For example, molecular features, which form in cool high photospheric layers, are not expected to be blueshifted.

Hydrodynamic simulations including the effect of the radiation field are necessary to better understand granulation. Two-dimensional, as well as three-dimensional radiative-hydrodynamic models have been computed for the Sun \citep[e.g.,][]{nordlund82,steffen91,stein98,robinson03,voegler04} as well as other stars \citep[e.g.,][]{nordlund90,asplund05:review,collet06,collet07,trampedach07}. Despite the fact that, due to computational limitations, these models describe only a small, hopefully representative, piece of the stellar envelope and have a limited spatial resolution, they have been remarkably successful at reproducing the observed properties of the solar \citep[e.g.,][]{dravins81,steffen91,asplund00:iron_shapes} and stellar granulation \citep[e.g.,][]{dravins90:disk_integrated,allende02}. These theory vs.~observation comparisons, however, have been performed for only a few stars and often using a limited set of spectral features. Thus, there is an urgent need for quantitative testing of the fundamental predictions of these hydrodynamic simulations in order to better understand stellar granulation and its impact on the observed spectra.

Of particular interest is the impact of granulation on the determination of fundamental parameters and chemical compositions of stars. With few exceptions, most of these studies use ``classical'' one-dimensional model atmospheres, which are calculated adopting plane parallel geometry, hydrostatic equilibrium, energy flux conservation, and local thermodynamic equilibrium (LTE), in addition to ignoring the effects of magnetic fields, rotation, and chromospheric activity \citep[e.g.,][]{kurucz79,kurucz93:cd13,gustafsson75,gustafsson08}. Although state-of-the-art hydrodynamic simulations share with classical models many of these basic assumptions, they take into account the atmospheric inhomogeneities associated with the granulation phenomenon.

Spectroscopic studies using classical model atmospheres have provided a wealth of useful information for general astrophysics and cosmology (see, e.g., the reviews by \citealt{wheeler89} and \citealt{mcwilliam97}) and it is, therefore, very important to check whether the abundances determined using classical methods are self-consistent. Exploring the impact of granulation (also referred to as ``3D effects'') on abundances and fundamental parameters derived using classical methods, in particular for K-dwarfs (see below), is therefore crucial to improve our knowledge of cosmochemistry.

Since the lifetimes of K-dwarfs ($\sim20$~Gyr) are larger than the age of the Galaxy \citep[$13.73\pm0.12$~Gyr,][]{hinshaw08}, no K-dwarf (and, evidently, no other cooler dwarf star) has evolved off the main sequence yet. Thus, as opposed to F and early G main-sequence stars, samples of K-dwarfs are free from biases in chemical evolution studies due to stellar death. Although M-dwarf and cooler star samples are also free from these biases, the analysis of their spectra is complicated by the abundance of molecules and condensates (dust, clouds), uncertain opacities and molecular equilibrium calculations, which make the model atmosphere computation difficult \citep[e.g.,][]{hauschildt99,burrows06} and the spectroscopic analysis arduous (ill-defined continuum, atomic lines blended with molecular bands). In addition, they are very faint and therefore difficult to observe at high resolution and signal-to-noise ratio ($S/N$) \citep[e.g.,][]{woolf05, bean06}.

K-dwarf spectra are also useful for radial velocity determinations due to the large number of spectral lines present. Currently, radial velocity uncertainties of a fraction of 1~km~s$^{-1}$ are considered reasonable for studies of kinematics and stellar abundances. Even with the highest $S/N$ and highest resolving power spectra, these uncertainties cannot be reduced, in an absolute sense, below a level of a few hundreds of meters per second due to systematic errors. One of these is the relativistic gravitational redshift, which can be, however, reasonably well estimated from stellar mass and radius determinations, at least for nearby stars \citep[e.g.,][]{dravins99}. The other important source of uncertainty is due to granulation. 

Typically, granulation effects introduce a radial velocity uncertainty on the order of several hundreds of meters per second, an error that depends on the lines used, given that wavelength shifts depend on the properties of the spectral line, in particular line strength. If both gravitational redshift and convective blueshift are properly taken into account, the uncertainty in the \textit{absolute} radial velocity of the standard stars, some of which are K-dwarfs \citep[e.g.,][]{stefanik99}, could be reduced dramatically to only a few tens of meters per second.

Standard stars with very accurate absolute radial velocities can be crucial for the interpretation of data from massive kinematic (and abundance) surveys such as RAVE \citep{steinmetz06}, GAIA \citep[e.g.,][]{perryman05}, APOGEE \citep{majewski07}, and SDSS/SEGUE \citep{refiorentin07}. In particular, constant radial velocity objects can replace comparison lamps as wavelength calibrators, if needed (for example, this has been proposed for the GAIA mission, where the observations of a group of ``well-behaved'' stars will be used to self-calibrate its very stable spectrograph; \citealt{katz04}). The determination of accurate radial velocities is also important for the detection of long term variability (e.g., unresolved wide binaries). Furthermore, the elimination of systematic errors from spectroscopically determined radial velocities is necessary to assess the reliability of non-spectroscopic methods of radial velocity determination, such as those involving astrometric measurements, for example via secular parallax/proper motion variations or the varying extent of moving clusters \citep{dravins99}. While these techniques were envisioned as early as the 1900's, their applicability has been limited by the accuracy of the astrometric data. Currently, for example with \textit{Hipparcos} \citep{perryman97} data, only the moving cluster method can in some cases achieve accuracies below 1~km\,s$^{-1}$ \citep{lindegren00,madsen02} but the advent of future space astrometry missions (in particular GAIA), which will improve the accuracy of stellar parallaxes and proper motions by at least two orders of magnitude, guarantees that these techniques will become very powerful and the results widely used.

\medskip

In this series of papers, we perform a detailed study of granulation in K-dwarfs and explore its impact on the observed spectra, in particular line shapes and wavelength shifts, as well as spectroscopic abundances and fundamental parameters. In this paper (Paper~I), we look for the basic signatures of granulation on very high quality spectra of a small sample of K-dwarfs. \citetalias{kdwarfs-p2} \citep{kdwarfs-p2} deals with hydrodynamic simulations of K-dwarf envelopes, and in \citetalias{kdwarfs-p3} \citep{kdwarfs-p3} we explore the impact of 3D effects on the determination of stellar abundances and fundamental parameters.

The data acquisition and processing are described in detail in this paper. Therefore, some sections are more relevant to \citetalias{kdwarfs-p2} or \citetalias{kdwarfs-p3} than to this paper. For example, the discussion about continuum normalization (Sect.~\ref{s:norm&merg}) will be important for our study of 3D effects on spectroscopic abundance determinations in \citetalias{kdwarfs-p3} while the details about the instrumental profile (Sect.~\ref{s:resolution}) will be crucial for our comparison of predicted to observed line profiles in \citetalias{kdwarfs-p2}.

\section{Detection of granulation signatures in K-dwarf spectra} \label{s:detection}

At least 5 independent points across a spectral line are required to define its bisector \citep[e.g.,][]{dravins87:observability}. Thus, for typical strong lines in K-dwarf spectra with a full width of about 10~km\,s$^{-1}$, we need at least 2~km\,s$^{-1}$ resolution, or $R=\lambda/\Delta\lambda=150,000$. In practice, however, the feasibility of measuring line bisectors also depends on the magnitude of the granulation effects and the level of noise. In K-dwarfs, granulation effects are very small, on the order of tens to a few hundreds of meters per second. Numerical tests show that only at very high resolving power ($R\gtrsim150,000$) and high signal-to-noise ratio ($S/N\gtrsim1,000$) the line bisectors associated with granulation are well defined. To reach such high $S/N$, however, very long exposure times are required for typical K-dwarfs (even the brightest, nearest ones), which makes the observations impractical for mid-sized telescopes. At $S/N\simeq300$ and $R\simeq200,000$, the line bisectors are noisy, but the overall shapes of the bisectors due to granulation are still detectable.

In addition to the limitations due to finite $S/N$, line blends, which are particularly important in K-dwarfs, can severely distort the line bisectors. One way to overcome the effect of blends is to group bisectors of lines of the same species and similar strength, given that they form in roughly the same photospheric layers and are, therefore, expected to experience similar granulation temperature and velocity fluctuations. In this context, wide spectral coverage observations are ideal because they allow us to select a large number of clean lines from each spectrum.

For the measurement of wavelength shifts, on the other hand, it is necessary to have, in addition to very well calibrated spectra, accurate laboratory measurements of rest wavelengths. Unfortunately, these are currently at the level of about 75~\ms (Sect.~\ref{s:wavshifts}), and our results are therefore subject to these uncertainties. The type of study presented here would therefore greatly benefit from efforts by laboratory physicists to improve these wavelength measurements.

\begin{table*}
\centering
\begin{tabular}{rcrccccl}\hline\hline
HIP & $V$ & $d$ & SpT & $M_V$ & $\teff$ & $\feh$ & Notes \\
       & mag & pc  &     &  mag  & K       & dex  &  \\
\hline
37279 & 0.3 & $3.50$ & F5 & 2.7 & 6677 & +0.08 & Procyon, $\alpha$ CMi, HD 61421, HR 2943, GJ 280 A \\
      & $\cdots$ & $\cdots$ & G2 & 4.8 & 5777 & +0.00 & skylight \\
96100 & 4.7 &  $5.77$ & K0 & 5.9 & 5218 & $-0.22$ & $\sigma$ Dra, HD 185144, HR 7462, GJ 764 \\
26779 & 6.2 & $12.24$ & K1 & 5.8 & 5150 &  +0.11 & HD 37394, HR 1925, GJ 211 \\
16537 & 3.7 &  $3.22$ & K2 & 6.2 & 5052 & $-0.08$ & $\epsilon$ Eri, HD 22049, HR 1084, GJ 144 \\
88601 & 4.0 &  $5.09$ & K0 & 5.5 & 5050 & $-0.04$ & 70 Oph, HD 165341, HR 6572, GJ 702 \\
64797 & 6.5 & $11.23$ & K2 & 6.0 & 4915 & $-0.15$ & HD 115404, GJ 505 A \\
37349 & 7.2 & $14.20$ & K2 & 6.4 & 4889 &  +0.00 & HD 61606, GJ 282 A \\
86400 & 6.5 & $10.71$ & K3 & 6.4 & 4833 & $-0.05$ & HD 160346, GJ 688 \\
114622 & 5.6 &  $6.53$ & K3 & 6.5 & 4743 &  +0.09 & HD 219134, HR 8832, GJ 892 \\
23311 & 6.2 &  $8.81$ & K3 & 6.5 & 4641 &  +0.26 & HD 32147, HR 1614, GJ 183 \\
\hline
\end{tabular}
\caption{Sample observed in this work. Basic data ($V$, SpT) were obtained from SIMBAD. Distances and absolute magnitudes are based on \textit{Hipparcos} parallaxes. The atmospheric parameters $\teff$ and $\feh$ are those determined by \cite{allende04:s4n} and have typical uncertainties of about 100~K and 0.1~dex, respectively.}
\label{t:sample}
\end{table*}

\section{Observations} \label{s:observations}

\subsection{Target selection} \label{s:selection}

The Spectroscopic Survey of Stars in the Solar Neighborhood (S$^4$N, \citealt{allende04:s4n}) contains all G and early K-dwarfs (K2V and earlier) within 15~pc. Cooler stars are also included in the survey but their sampling is not complete. We used this catalog, and the stellar parameters derived there, to select our targets by constraining their effective temperatures to the range 4600\,K\,$\lesssim\teff\lesssim$\,5250\,K and metallicities to $-0.25\lesssim\feh\lesssim+0.25$, in addition to their observability from McDonald Observatory. Nine K-type dwarf stars were selected and are listed in Table~\ref{t:sample}.

We also obtained spectra of Procyon and the day skylight, the latter as a proxy for the solar spectrum, to check the accuracy of our data reduction procedures, given that these stars have very high quality spectra available from previous studies \citep[e.g.,][]{griffin96,kurucz84}. Note, however, that the skylight spectrum obtained in this manner may be slightly different than the real solar spectrum because the sky fills the spectrograph slit completely, therefore making the data reduction different from that of a point source, and it is known that aerosol and Rayleigh-Brillouin scattering affect slightly the observed line shapes and strengths \citep{gray00}.

\subsection{Observations and data reduction} \label{s:obsanddatarx}

The data for our sample stars were acquired in 5 observing runs between February 2006 and May 2007. The spectra were obtained with the 2dcoud\'e spectrograph \citep{tull95} on the 2.7\,m Harlan J. Smith Telescope at McDonald Observatory, using the cs21 mode (grating e2, focus f1), which is expected to deliver a spectral resolution of $R\simeq210,000$ (see, however, Sect.~\ref{s:resolution}). In the cs21 mode, a single exposure results in an echelle spectrum of about 20 orders, each of them covering approximately 20\,\AA. By rotating the spectrograph grating to 7 different positions (hereafter referred to as ``instrumental setups''), full coverage from 5580 to 7900\,\AA\ was achieved.

Previous observational studies of granulation in K-dwarfs  \citep[e.g.,][]{dravins87:line_asymmetries,gray82,gray05} have used very high quality data but with an otherwise limited spectral coverage, sometimes involving the analysis of only one spectral feature. Although the quality of our observations is similar or slightly superior to that of these pioneering studies, our wavelength coverage surpasses all of them.

A typical single exposure of 20 minutes for a $V\simeq6$ star resulted in $S/N\simeq100$ under good observing conditions. To reach our required signal-to-noise ratio ($S/N\gtrsim300$), several single exposures of the same setup were co-added before merging them into a single spectrum. Sects. \ref{s:shift} and \ref{s:norm&merg} explain the details of the co-adding and merging procedures.

\begin{figure*}
\centering
\includegraphics[width=18.5cm,bb=70 375 680 740]{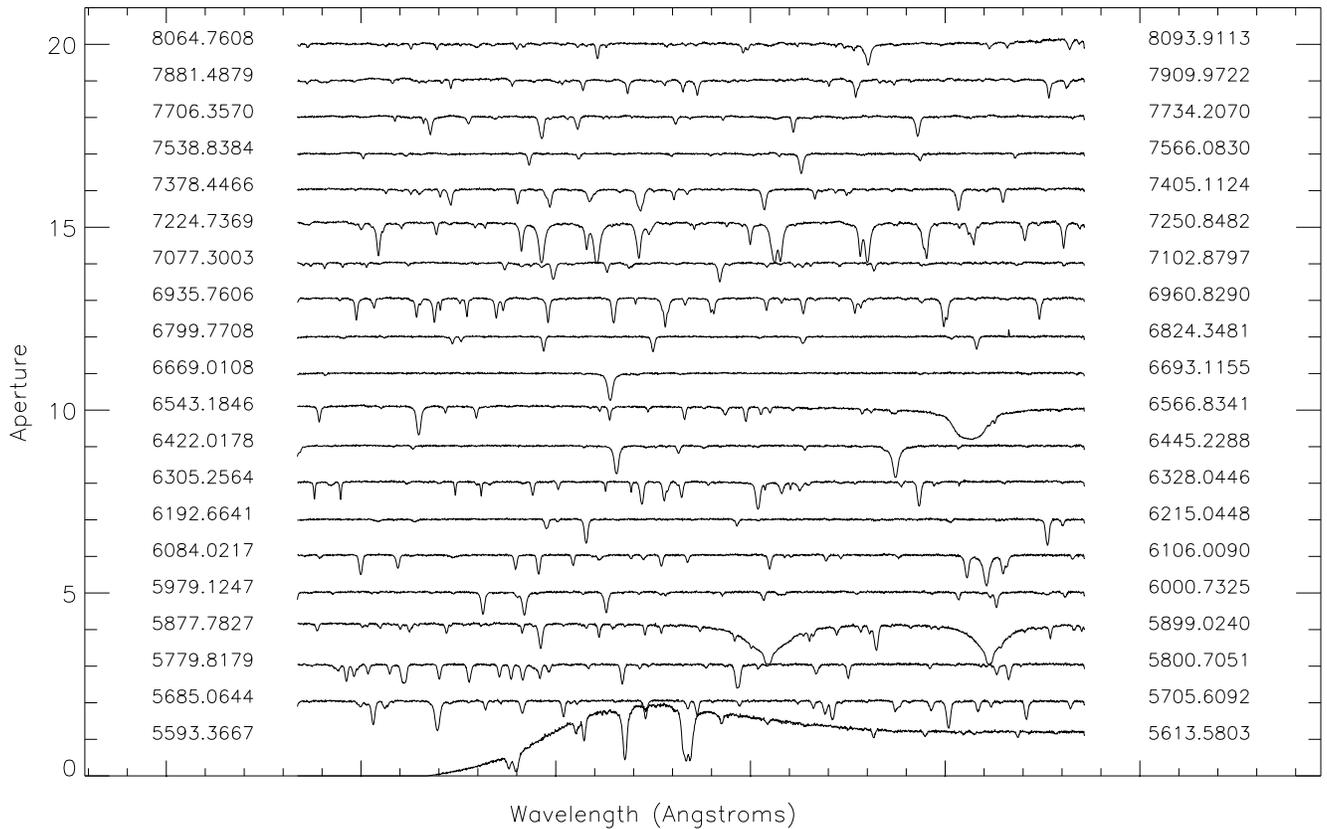}
\caption{Sample spectrum from our 2dcoud\'e--cs21 observations for HIP~96100. Only one instrumental setup is shown.}
\label{f:samplespec}
\end{figure*}

Data reduction (bad pixel removal, overscan correction, flat-fielding, and scattered light correction) was performed using the standard IRAF\footnote{IRAF is distributed by the National Optical Astronomy Observatories, which are operated by the Association of Universities for Research in Astronomy, Inc., under cooperative agreement with the National Science Foundation -- {\tt http://iraf.noao.edu}} {\tt echelle} package. Comparison lamp (ThAr) spectra were obtained approximately every hour to ensure an accurate wavelength calibration ($5\times10^{-4}$\,\AA\ of RMS scatter on average). Instead of linearizing the wavelength solutions (i.e., resampling each order to a wavelength scale of constant step), the actual pixel-to-wavelength relations were preserved to minimize interpolation errors in the observed counts. Fig.~\ref{f:samplespec} shows an example of a reduced spectrum.

In addition to the 2dcoud\'e data, we obtained spectra of our sample stars using the High Resolution Spectrograph \citep[HRS;][]{tull98,hill06} on the 9.2\,m Hobby-Eberly Telescope (HET) for wavelength calibration purposes (details are given in Sect.~\ref{s:norm&merg}). This instrument is capable of delivering $R\simeq120,000$ spectra with complete wavelength coverage from 6000 to 7900\,\AA\ (red chip) and from 4100 to 5900\,\AA\ (blue chip) in one single exposure. The process of data reduction was similar to that applied to the 2dcoud\'e data. For these spectra, the accuracy of the wavelength solutions is about $10^{-3}$\,\AA\ for the red chip and $5\times10^{-4}$\,\AA\ for the blue chip. Exposure times were set to reach $S/N\simeq750$ at the center of the red chip. The procedures described in the next section were also applied to these data.

\subsection{Shifting and co-adding: increasing the $S/N$ ratios} \label{s:shift}

As explained above, several individual exposures of each setup were required to reach high signal-to-noise ratios. Co-adding them is not trivial because even after correcting for the Earth's motion, wavelength shifts due to the intrinsic radial velocity of the star (e.g., due to binarity, variability, and presence of planets) or instrumental and environmental effects (the 2dcoud\'e spectrograph is not inside a vacuum chamber) are present between exposures. These shifts are on the order of tens to hundreds of meters per second and, therefore, of the same magnitude as the effects we are trying to resolve. Consequently, these shifts must be taken into account before co-adding in order to keep the spectra useful for granulation studies.

We determined radial velocity shifts between pairs of spectra of the same object by cross-correlation in the Fourier space, after rebinning the spectra to have a constant step in $\log\lambda$. Orders severely affected by telluric lines (more than 50\% of the $\lambda$ range) were discarded. Our cross-correlation calculations were made using the IDL {\tt xc} code by \cite{allende07} and not the popular IRAF {\tt fxcor} task because the latter requires the spectra to be linearized, a procedure that we avoided (see Sect.~\ref{s:obsanddatarx}).

Determining the uncertainty of our cross-correlation procedure is a difficult task. Monte~Carlo simulations using both synthetic spectra and an observed high $S/N$ spectrum predicted errors of only about 4~m~s$^{-1}$ for $R=200,000$ and $S/N=100$. This procedure leads to optimistic error estimates because it does not take into account other effects such as pixel shifts that are non-linear in velocity and order-to-order scatter due to instrumental distortions, the fact that spectral lines are not symmetric or at their rest wavelengths due to granulation effects, incompleteness or uncertainties in the atomic data used for the calculation of the synthetic spectra, telluric lines, etc. Instead, we estimated the error from the order-to-order scatter only, for which we found no strong correlation with wavelength. Thus we find that our relative radial velocities have a mean error of about 14~m\,s$^{-1}$.

The standard approach to the determination of radial velocity shifts between spectra is to cross-correlate them with a reference spectrum of known radial velocity. In our case, we do not have access to a reference spectrum of higher quality than the data (e.g., high resolution synthetic spectra including line asymmetries over the whole spectral range of our observations) and therefore we need to adopt a different approach. In this work, the number of spectra co-added for a single faint object in a given setup was typically between 5 and 10. Instead of cross-correlating each of them with a single reference spectrum, we cross-correlated all of them with each other. We then used all the measurements to refine the shifts of all spectra with respect to the first one of the list employing ``self-improvement'' \citep{allende07}. Using this technique, the mean error in the relative radial velocities was reduced by about 15\%.

After correcting for the radial velocity shifts, the spectra were coadded using an algorithm that minimizes interpolation errors in the observed fluxes. When coadding, each original pixel has to be interpolated to some common wavelength scale. One may be tempted to use the wavelength solution of one of the spectra and interpolate the rest to that dispersion at the risk of introducing large interpolation errors. Instead, we determined, for each set of spectra to be coadded, a wavelength solution such that the flux interpolation distances are minimized.

The pixel-to-wavelength relation of the first spectrum was used as a starting reference. The mean difference between adjacent pixels, $\delta\lambda$, was then determined. For each pixel in the reference spectrum, we defined a box extending to $\pm \delta\lambda/2$ around its wavelength. Next, we looked at the rest of the spectra and found all pixels with wavelengths within that box. The average wavelength of all those pixels was then adopted for that particular position. In this way, the mean interpolation distance was reduced by a factor of 2, compared to the case in which the spectra are interpolated to the dispersion of the first (or any other reference) spectrum. The interpolation distance for the linear case (i.e., rebinning the spectra to a common constant step dispersion) is substantially larger (a typical factor of 5 to 10) but, as explained above, we avoided this type of interpolation.

\subsection{Normalization and ``merging''} \label{s:norm&merg}

As explained in Sect.~\ref{s:selection}, all of our sample stars are included in the S$^4$N survey of \cite{allende04:s4n}, which provides atlases of spectra covering the wavelength range from 362 to 921~nm with a resolving power $R\simeq60,000$. The continuum normalization of these data is superb; not only were the blaze shapes removed by fitting high order polynomials to the upper envelopes of the observed fluxes in each order, but also the smooth variation of the blaze shapes in the direction perpendicular to the dispersion was taken into account, thus removing the instrumental response with a very careful two-dimensional modeling, as described in \cite{barklem02}. This normalization was essentially inherited by our spectra using the procedure described below.

Each order of our coadded cs21 spectra (one instrumental setup at a time) was divided by its corresponding cs23 piece from the S$^4$N data set, after lowering the resolution of the cs21 data to match that of the cs23 data, correcting for radial velocity shifts, and rebinning to a common wavelength scale. In principle, the result should be a smooth function corresponding to the shape of the continuum (the upper ``envelope'') of the cs21 observations. However, due to the finite $S/N$ values and temporal variations of the strengths of some lines (in particular the telluric ones), this envelope had to be smoothed out using a median filter of width equal to 100 pixels (about 2\,\AA). The cs21 data were then divided by this envelope in each order, for each instrumental setup.

\begin{figure*}
\centering
\includegraphics[width=18.6cm,bb=40 180 580 435]{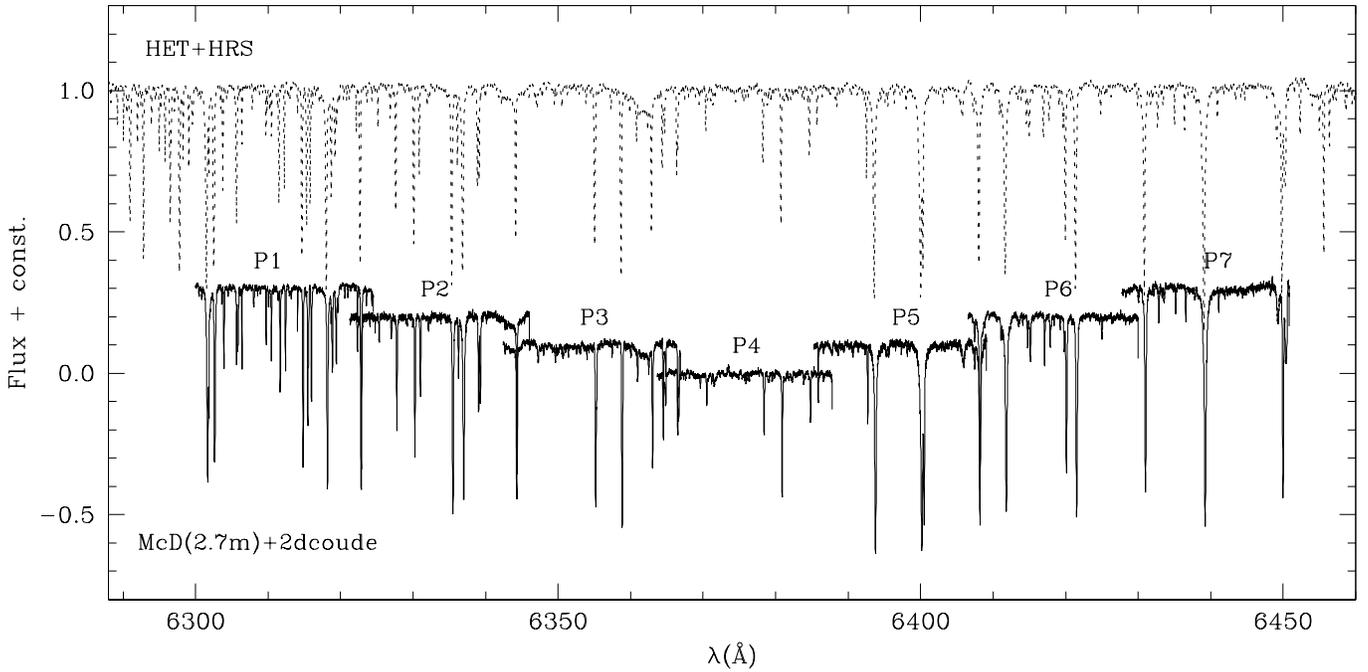}
\caption{Merging of spectra from different instrumental setups (P1-7). Only one order per setup is shown here; note that each setup has about 20 orders in different wavelength regions (see Fig.~\ref{f:samplespec}). The fluxes have been shifted arbitrarily in the vertical direction for clarity. The cs21 pieces (P1-7) have been aligned with the HET+HRS spectrum by means of cross-correlation and are ready to be merged.}
\label{f:hetpuzzle}
\end{figure*}

The coadded and normalized cs21 data (7 setups for each object) were finally merged using an HET spectrum of the same object as a radial velocity template. Note that the spectra for each setup can have a different radial velocity, since we only removed relative, and not absolute, Doppler shifts (Sect.~\ref{s:shift}). Therefore, each order of a cs21 setup was cross-correlated with its corresponding piece in the HET-HRS spectrum, after lowering the resolution of the former to match the $R\sim120,000$ of the HRS, to determine the radial velocity needed to align the cs21 spectrum with the HET-HRS spectrum, which has a unique wavelength solution. The robust (iterated and weighted) average shift of all orders for each setup was adopted to correct the cs21 data, thus removing inaccurate shifts from orders affected by telluric lines or instrumental defects. In the regions of overlap, a weighted mean for the fluxes was calculated after adopting the common robust rebinning explained in Sect.~\ref{s:shift}. Fig.~\ref{f:hetpuzzle} illustrates this step of the data processing.

On average, the order-to-order scatter of the radial velocities used for the merging of setups was 70~m~s$^{-1}$. This error is larger than the 12~m~s$^{-1}$ error quoted in Sect.~\ref{s:shift} for the order-to-order scatter of the cs21 data alone, in part due to the lower resolving power of HRS, but also probably because of distortions in the wavelength-to-pixel mapping of HRS. Note that the regions of overlap of the different cs21 setups are small compared to the wavelength coverage of each order. Only within the regions of overlap, spectral line shapes are affected by this large merging error of 70~m~s$^{-1}$. The majority of spectral lines (approximately 80\% of them), however, are not in the regions of overlap and the wavelength mapping within a spectral line is still accurate at the 12~m~s$^{-1}$ level. On the other hand, in the final merged spectrum, the absolute wavelength scale of all setups and orders combined is uncertain to about 70~m~s$^{-1}$. This 70~m~s$^{-1}$ uncertainty will have an effect on our measurements of wavelength shifts (Sect.~\ref{s:wavshifts}) but not on the line profiles or their bisectors.

\subsection{Resolving power and instrumental profile} \label{s:resolution}

To calculate the resolving power of 2dcoud\'e-cs21, we used the ThAr exposures. We measured the full width at half maximum (FWHM) of individual Th ($f_\mathrm{Th}$) and Ar ($f_\mathrm{Ar}$) lines using the line identifications of \cite{murphy07}. Note that the Th lines are sharper than the Ar lines due to their different atomic weights, which makes thermal broadening more important for the latter. The thermal broadening ($\sigma$) is inversely proportional to the square root of the atomic weight $A$; thus
\begin{equation} \frac{\sigma_\mathrm{Th}}{\sigma_\mathrm{Ar}}=\left(\frac{A_\mathrm{Ar}}{A_\mathrm{Th}}\right)^{1/2}=0.415.
\end{equation}
If the instrumental FWHM is $f$, then
\begin{equation}
f^2+\sigma_\mathrm{Th}^2=f_\mathrm{Th}^2 \label{eq:fth}
\end{equation}
\begin{equation}
f^2+\sigma_\mathrm{Ar}^2=f_\mathrm{Ar}^2 \label{eq:far}
\end{equation}
from which we obtain an expression for the instrumental FWHM that is independent of temperature:
\begin{equation}
f^2=\frac{f_\mathrm{Th}^2-0.17f_\mathrm{Ar}^2}{0.83}\ .\label{eq:fwhm}
\end{equation}
The resolving power is then $\lambda/f$.

\begin{figure}
\centering
\includegraphics[bb=80 390 340 640,width=8cm]{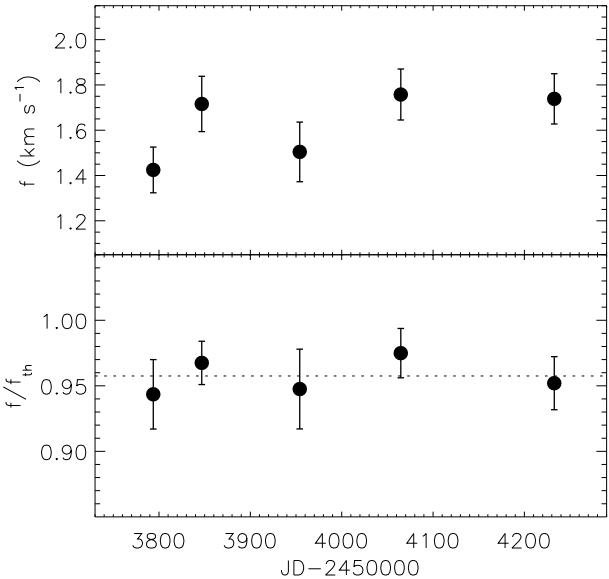}
\caption{Upper panel: Average FWHM of the instrumental profile for each of our observing runs. The 2dcoud\'e spectrograph (cs21 mode) is expected to deliver $f\simeq1.4$~km\,s$^{-1}$ at all times; the differences shown here reveal that the spectrograph has suffered from focus degradation. Lower panel: Average ratio of instrumental to Th line FWHM values for each of our observing runs. The constancy of this ratio suggests that the differences seen in the upper panel of this figure are not related to temperature effects on the ThAr lamp.}
\label{f:fwhm_mean}
\end{figure}

The measured FWHM values were found to vary significantly with time, even within a given run of a few days. From one run to another (months apart), the change in the average FWHM values was significant, as shown in the upper panel of Fig.~\ref{f:fwhm_mean}. Temperature broadening effects alone cannot explain the diversity of values seen in Fig.~\ref{f:fwhm_mean} given that less than 5\% of the broadening of Th lines is thermal (see lower panel of Fig.~\ref{f:fwhm_mean}). Considering changes of 10\% in the temperature of the ThAr lamp (which was actually well controlled), the total broadening would be affected by only $0.05\times\sqrt{0.1}$=1.6\% or less, yet in Fig.~\ref{f:fwhm_mean} (upper panel) we see variations of up to 20\%. The minimum and maximum values of the resolving power of our observations are about 160,000 and 210,000, respectively.

\begin{figure}
\centering
\includegraphics[bb=70 360 280 863,width=8cm]{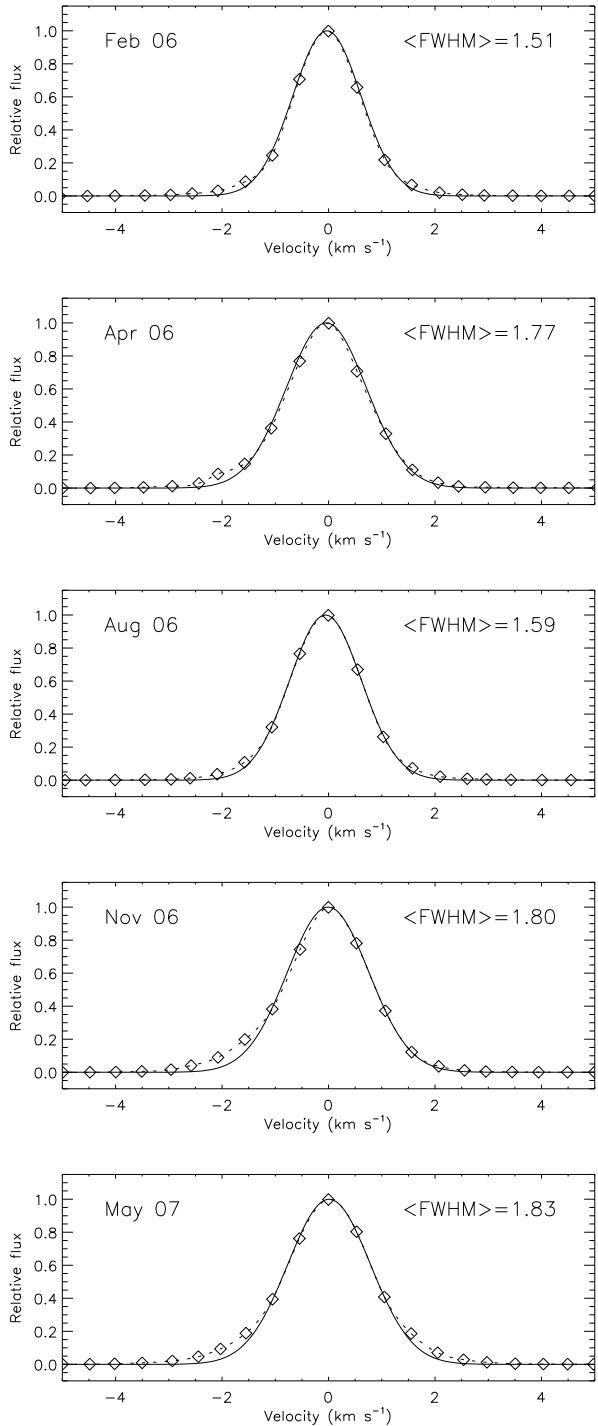}
\caption{The average instrumental profile for each of our observing runs is shown with diamonds connected by dotted lines. The solid lines correspond to Gaussian profiles of FWHM identical to the average FWHM of the instrumental profiles, which are given in the upper right corner of each panel in km~s$^{-1}$.}
\label{f:psf_final4}
\end{figure}

Note that in Eqs.~\ref{eq:fth} and \ref{eq:far}, as well as in the discussion given above, it was assumed that the instrumental profile of the spectrograph was Gaussian. Although the actual ThAr line profiles are slightly asymmetric, Gaussian fits made reasonably good approximations, as shown in Fig.~\ref{f:psf_final4}, where the \textit{average} Th line profiles for each observing run are shown. They were obtained from the individual Th line profiles, normalized to 1 at their maximum before averaging in bins of 0.5~km\,s$^{-1}$, which corresponds to about one-third of the width of the resolution element. The departure from Gaussianity of the real instrumental profile shown in Fig.~\ref{f:psf_final4} is very small. The most prominent non-Gaussian feature is observed on the blue wing of the instrumental profile (near $-2$~km~s$^{-1}$) and it amounts to less than 10\,\%. In fact, when the instrumental profile was sharpest (in February 2006), the maximum difference between the real profile and the Gaussian profile was only about 5\%. A similar bump on the red wing is observed only for the May 2007 observations, which correspond to the most degraded (wider) instrumental profile of our observations. This information will be crucial for the comparison of observed line profiles with those predicted by hydrodynamic model atmospheres in \citetalias{kdwarfs-p2}.

\section{Measurement of the granulation signatures} \label{s:measurement}

The spectra of K-dwarfs, in particular those with near solar and super-solar metallicities, are very rich in absorption lines. For spectroscopic studies of granulation, iron is an ideal element; it has a large atomic mass, which implies low thermal broadening and therefore a high sensitivity to non-thermal fields, one isotope is much more abundant than the others, thus reducing the impact of asymmetries due to fine structure, it is abundant, and its atomic structure is very complex, allowing a very large number of transitions. Furthermore, contrary to most other heavy elements, reliable laboratory data are available for iron. In this paper, we deal with the detection of the granulation signatures using \fei\ lines exclusively. Although the granulation effects are expected to be stronger in \feii\ lines owing to their larger formation depth, few \feii\ features are available in our spectra and almost all of them are severely affected by blends.

\subsection{Line bisectors} \label{s:linebis}

We measured the bisectors of a large number of spectral lines listed in the \cite{nave94} multiplet table for Fe~\textsc{i}. In our available spectral range ($5580<\lambda<7800$\,\AA), \citeauthor{nave94} list about 900 lines, most of which are weak or blended. For each spectrum, we examined the neighboring $\pm1$\,\AA\ of all $\sim900$ Fe~\textsc{i} lines mentioned above and attempted to measure their line bisectors. Very weak lines with minimum normalized flux values above 0.98 were discarded because their bisectors are very sensitive to blends and noise. Spectral lines in regions where the local $S/N$ was below 200 (as estimated from the total number of counts in the local continuum) were also discarded, except for HIP~37349, which has $S/N\simeq150$ everywhere.

The line bisectors were measured by determining iso-flux wavelength (or velocity) points on each wing of the line, resampling the original profiles using cubic spline interpolations. The error bars were obtained by propagating the observational errors of our spectra, as quantified by the local continuum $S/N$ (see \citealt{gray83} for details). Hereafter, the line bisectors are measured with respect to the line core wavelength. The latter measurements are discussed separately in Sect.~\ref{s:wavshifts}.

Many of the Fe~\textsc{i} lines given in the \citeauthor{nave94} catalog are blended in the observed stellar spectra. In many cases the blends will be clearly distinguishable by visual inspection and a line selection could be made in this manner. However, even weak line blends can significantly distort the shape of the line bisectors, and, given that the line asymmetries are very small in K-dwarfs, these blends should be avoided using a different approach. Instead of selecting subjectively spectral lines based on visual inspection of the line itself or its bisector, we adopted an automated scheme in which the \textit{smoothness} of the line-bisector is used as evidence that the line is not blended. For each bisector measured, we computed the slope $d(\Delta\mathrm{v})/df_n$, with $\Delta\mathrm{v}$ being the abscissa values of the line bisector and $f_n$ their corresponding normalized fluxes. Bisectors for which $d(\Delta\mathrm{v})/df_n>4,000$~m~s$^{-1}$ per normalized flux unit were discarded.\footnote{On average, we find that bisectors of strong lines have $d(\Delta\mathrm{v})/df_n\simeq150$~\ms per normalized flux unit and therefore this criterion is a conservative but robust one.}  Of course, lines blended with features that share the same core wavelength will survive this criterion but their number will be small compared to blends on the line wings and, furthermore, even subjective criteria will have difficulty disentangling them from true clean lines. About 40\% of the lines listed by \citeauthor{nave94} survived our selection criterion.

\begin{figure*}
\includegraphics[bb=80 395 600 540,width=16cm]{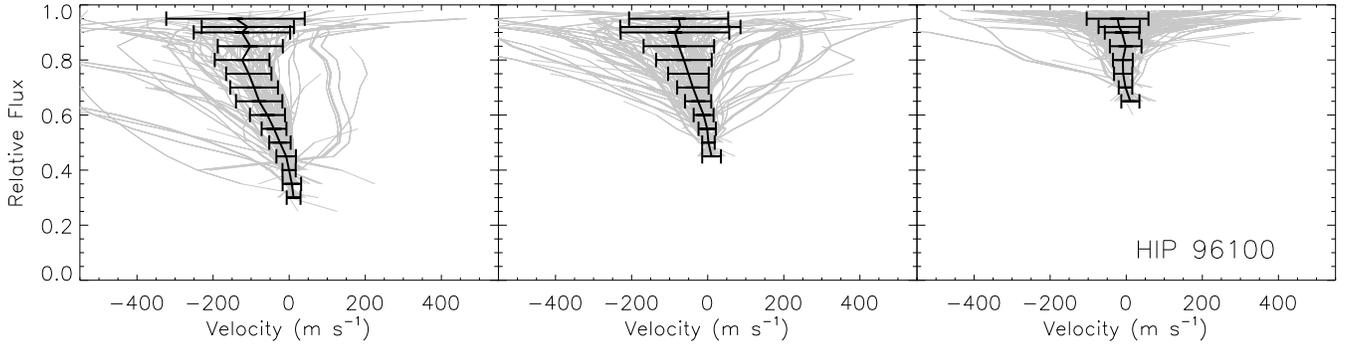}
\caption{Mean bisectors for HIP~96100 (black solid lines with error bars) obtained from individual line bisectors (gray lines). The error bars correspond to a $1\sigma$ scatter.}
\label{f:meanbissample}
\end{figure*}

\begin{figure*}
\includegraphics[width=17.2cm,bb=93 380 450 680]{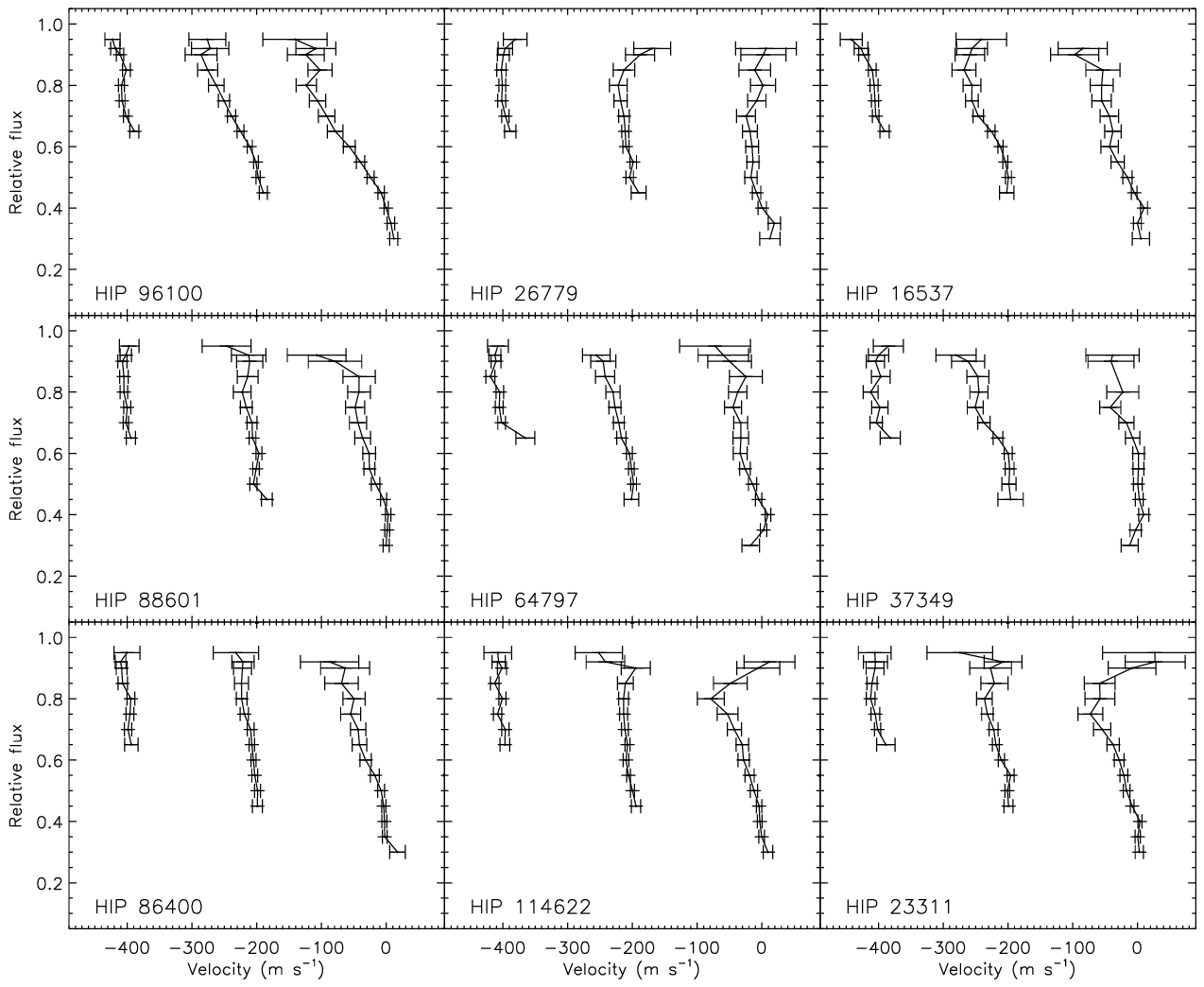}
\caption{Mean line bisectors for our sample of K-dwarfs, measured with respect to their core wavelengths. Shifts of $-200$ and $-400$~m~s$^{-1}$ have been applied to the weaker lines for clarity. The error bars correspond to the standard error of the line-to-line scatter, i.e., $\sigma/\sqrt{n_l}$, where $\sigma$ is the standard deviation and $n_l$ the number of lines. Bisector points for which $\sigma>200$~\ms have been discarded (these often appear near the continuum).}
\label{f:mbisk}
\end{figure*}

Even after this selection of smooth bisectors, the level of noise was relatively high, as expected for the typical signal-to-noise ratios of our observations. Since the shape of the line bisector depends primarily on the line strength, the noise can be reduced by averaging bisectors of spectral lines of similar strength. This is shown in Fig.~\ref{f:meanbissample} for the case of the star HIP~96100. Although a wide range of bisector shapes is apparent in Fig.~\ref{f:meanbissample}, we applied robust statistics to obtain the mean bisectors, iteratively discarding points that disagree from the mean values by more than 2.5$\sigma$, which makes our final results reliable. Similar mean \fei\ line bisectors were computed for all the K-dwarfs in our sample and are shown in Fig.~\ref{f:mbisk}.

It is reasonable, although not strictly correct, to assume that the line bisectors of lines of similar strength are identical. In that case, the standard deviation ($\sigma$) is simply a measure of the noise and the actual measurement error of the mean line bisectors should be given as the standard error $(\sigma/\sqrt{n_l})$ of the $n_l$ line bisectors used in each group. In Fig.~\ref{f:mbisk}, the error bars correspond to the standard error.

The mean line bisectors for HIP~96100 have the largest span\footnote{We define the ``span'' as the difference in velocity between the reddest and bluest points of the line bisector.}
among our sample stars. There are three reasons that can explain this. First, the star has the hottest effective temperature in the sample, second, its metallicity is lower than solar ($\feh=-0.22$), and, finally, its projected rotational velocity is the lowest among our sample stars ($V\sin i\simeq0.8$~km~s$^{-1}$, see Appendix~\ref{s:vsini}). The granulation effects, and in fact the absolute magnitude of the inhomogeneities, are predicted to increase with effective temperature and also with decreasing metallicity, although probably the latter effect is significantly smaller \citep[e.g.,][]{allende99}, while low projected rotational velocities ($V\sin i<1$~km~s$^{-1}$) do not alter the shape of the line bisectors significantly, as will be shown in \citetalias{kdwarfs-p2}. The dependence on effective temperature has in fact been observed in real stars \citep[e.g.,][]{gray:book}.

In this context, it is interesting to see the case of the next few hotter stars, HIP~26779, HIP~16537, HIP~88601, HIP~64797, and HIP~37349, which do not show bisectors with large spans. In fact, in some cases they are smaller than those of the coolest star in our sample (HIP~23311). The reason for this is probably their large $V\sin i$ values (approximately between 2 and 3~km~s$^{-1}$, see Appendix~\ref{s:vsini}), which reduces the span of the line bisectors \citepalias{kdwarfs-p2}. We note here that the data for HIP~64797 and HIP~37349 are noisier than the average. Most of the observations for HIP~64797 were made under poor weather conditions while the spectrum of HIP~37349 has only $S/N\sim150$.

In addition to the stellar rotation effect in the 5 stars mentioned above, it is also interesting to note that they show high levels of chromospheric activity (with the possible exception of HIP~88601), as evidenced by their Ca~\textsc{ii} H and K line profiles (Appendix~\ref{s:activity}). Chromospheric effects on spectral line formation have not yet been studied in enough detail to determine quantitatively their impact on our line bisector measurements. Nevertheless, the fact that the signatures of granulation seem to be slightly different in the most active stars of our sample suggests a correlation. An active chromosphere may have an impact on the line shapes and intensities in several ways. Emission may fill the cores of some lines, as it is obvious in the H and K lines, but it may be also present to a lesser extent in weaker lines. Zeeman splitting could be measurable due to the presence of strong magnetic fields  \citep[e.g.,][]{gray84:zeeman,borrero08}, while magnetic pressure may have an impact on the photospheric structure. We can conclude that the line bisectors are significantly affected by the presence of strong magnetic fields and/or stellar activity, although the magnitude and impact of these effects are largely unknown for stars other than the Sun, where this effect has been documented in some detail \citep[e.g.,][]{balasubramaniam02,schlichenmaier04}.

HIP~86400 is our reference star because it has parameters very close to those adopted in the calculation of the 3D model atmosphere that we will study in \citetalias{kdwarfs-p2}. While the mean bisector for the weak lines shows no significant asymmetry (a span of only about 15~\ms), the strongest lines show bisectors that span about 75~\ms on average.

The coolest stars in our sample, HIP~114622 and HIP~23311, both have relatively low $V\sin i$ ($\lesssim2$~km~s$^{-1}$, Appendix~\ref{s:vsini}). Their spectra are richer in spectral lines compared to those of our other sample stars and this is perhaps the reason why their mean bisectors show large deviations from the expected behavior near the continuum. In fact, the bisectors approach zero velocity there, which points to the influence of numerous blends randomly distributed on both sides of the spectral lines. Nonetheless, far from the continuum, their line bisectors are similar to those of the reference star HIP~86400. Therefore, it is likely that the granulation contrast does not change significantly in dwarf stars of effective temperature between 4600 and 4800~K.

\begin{figure}
\includegraphics[width=7.6cm,bb=95 385 250 543]{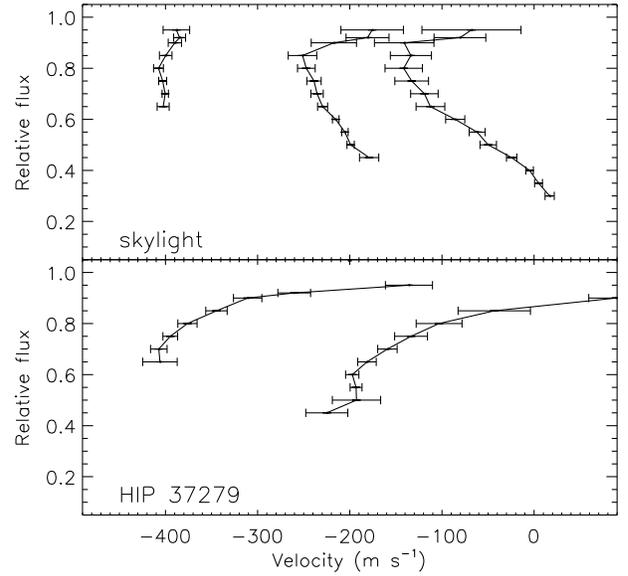}
\caption{As in Fig.~\ref{f:mbisk} for our observations of the Sun and Procyon.}
\label{f:mbissunp}
\end{figure}

To check that our measurements of line asymmetries are not affected by observational or data processing artifacts, we obtained mean line bisectors in our skylight and Procyon spectra as well (Fig.~\ref{f:mbissunp}). The mean bisectors of the skylight are much larger than those of our reference star. Note, however, that our skylight observations are significantly different than stellar ones. Solar line bisectors have been determined using the actual solar spectrum \citep[e.g.,][]{kurucz84} by several authors \citep[e.g.,][]{dravins81,asplund00:iron_shapes,allende02}, who show that their amplitudes are about twice as large as those shown in Fig.~\ref{f:mbissunp}, although this is in part due to the fact that their observations have higher spectral resolution (for example, the \citeauthor{kurucz84} atlas has a resolving power of 500,000). The mean bisectors measured in our Procyon spectrum, however, are in excellent agreement with those measured elsewhere \citep[e.g.,][]{dravins87:line_asymmetries,gray89:reversed,allende02}. Note that the metallic lines in the spectrum of this hotter star are, in general, weaker than those observed in K-dwarfs and, in our spectral range, very few strong lines are available. This is the reason why Fig.~\ref{f:mbissunp} does not show the strongest mean line bisector for Procyon.

\subsection{Wavelength shifts} \label{s:wavshifts}

The second fundamental test of granulation signatures in stellar spectra involves measuring core wavelength shifts. A large convective blueshift is expected to occur for the weakest lines, which are formed in deep layers and see strong granulation contrasts, while the cores of strong lines are much less affected given that they form in the highest photospheric layers, where the correlation between intensity and velocity fields is weak. We should therefore look at the relation between line shift and line strength (quantified, for example, by the line equivalent width) to determine the magnitude of the granulation effects and compare them to the model predictions.

Unfortunately, there are large uncertainties in the determination of the line shifts from our observed spectra, some of which are independent of the quality of our data. For example, very accurate laboratory wavelengths for large numbers of spectral features are necessary. The catalog by \cite{nave94} is perhaps one of the most accurate and complete line lists currently available for this purpose. Nevertheless, the wavelengths measured by them, in the best cases, have uncertainties of about 75~m~s$^{-1}$.

\begin{figure*}
\includegraphics[bb=100 380 480 680,width=17.7cm]{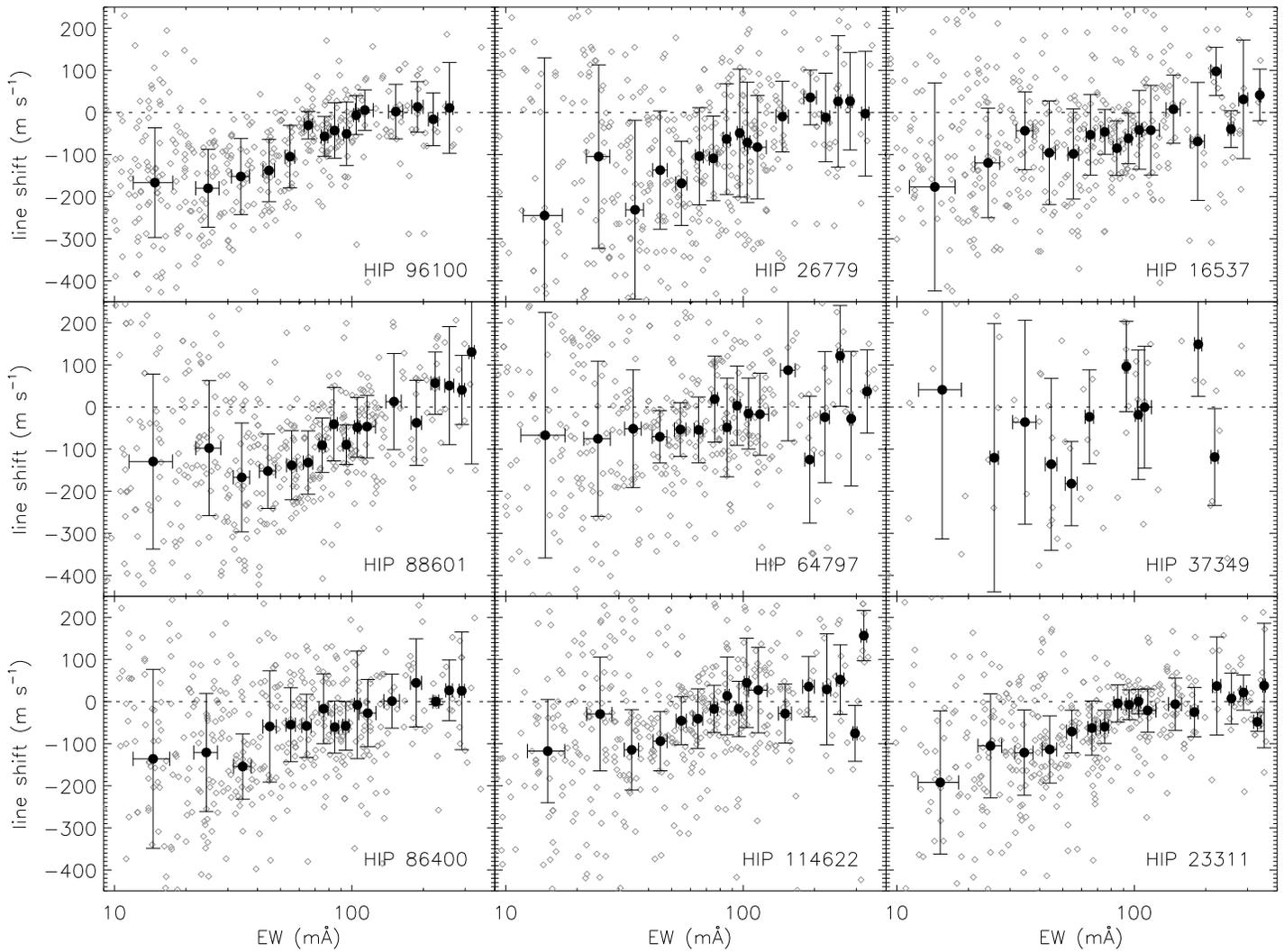}
\caption{Core wavelength shifts measured in the spectra of our sample stars as a function of line equivalent width (small diamonds). The filled circles with error bars represent average values in bins of equivalent width. The zero point of this relation has been set using the strongest lines, those that define the plateau.}
\label{f:lshmk}
\end{figure*}

We measured the core wavelengths of as many Fe~\textsc{i} lines as possible included in the \cite{nave94} catalog and restricted our measurements to those lines that have wavelengths measured with uncertainties lower than about 75~m~s$^{-1}$. The core wavelengths were obtained by fitting fourth order polynomials to the 11 points closest to the flux minimum. A more detailed description of the method we adopted to determine core wavelengths and their errors is given by \cite{allende02:shifts}.

Since the wavelength shifts in the solar spectrum have been measured with accuracy \citep[e.g.,][]{allende98} and the convective blueshift vs.~equivalent width relation is well defined there, we used our results for the skylight spectrum to perform a line selection. Only those lines for which the convective blueshifts were between $2\sigma$ of the mean line shift vs.~equivalent width trend were used. Statistically, this selection reduces the impact of line blends and misidentifications.

Our measured line shifts could be systematically displaced with respect to their absolute values due to the gravitational redshift and the uncertainty in the absolute stellar radial velocity. Thus, we adopted an arbitrary zero point using the core wavelength shifts of the strongest lines. Basically, the mean line shift of the strongest lines was set to zero (see below for details).

\begin{figure}
\includegraphics[bb=90 380 250 595,width=6.5cm]{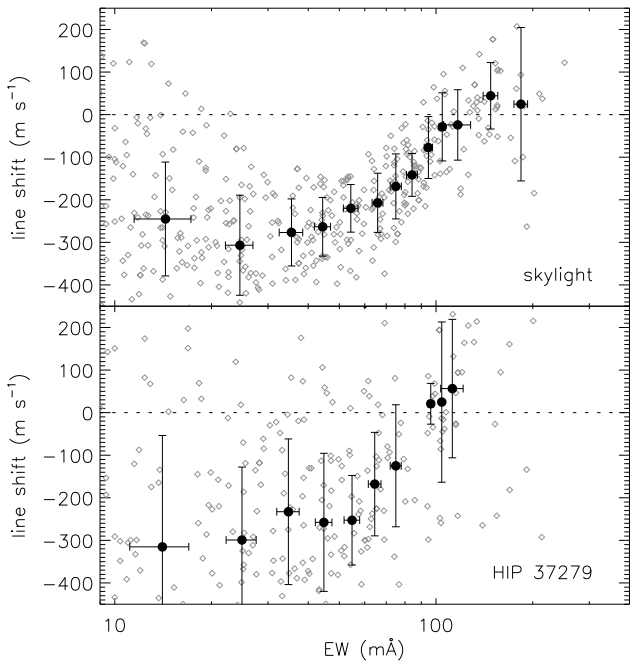}
\caption{As in Fig.~\ref{f:lshmk} for our observations of the Sun and Procyon.}
\label{f:lshmsunp}
\end{figure}

The line shift vs.~equivalent width relations for our sample stars are shown in Figs.~\ref{f:lshmk} and \ref{f:lshmsunp}. In each panel, we show the individual line shifts as well as averaged values in bins of equivalent width (the filled circles with error bars) to allow a better visualization given the large scatter observed. The bins have a width of 10~m\AA\ below $EW=100$\,m\AA\ and 35~m\AA\ above it. When the number of data points per bin is lower than 3, the bin-averaged values are not shown.

The basic signature of granulation is evident in all cases. The strongest lines show the smallest blueshifts while weaker lines are significantly blueshifted. Furthermore, the slope in the line shift vs.~equivalent width relation is larger for the skylight compared to the K-dwarf observations and even larger for Procyon. Although the scatter for the latter is large and there is not enough line strength coverage to draw in reasonably well the actual form and zero point of the trend (Fig.~\ref{f:lshmsunp}), this has been established by previous studies \citep[e.g.,][]{allende02}.

In several cases (e.g., HIP~96100, HIP 26779, HIP 86400, HIP 23311) it is clear that the trend is nearly linear for the weak lines but it reaches a saturation point at about 100~m\AA, from where it becomes a \textit{plateau}. The detection of this plateau is of utmost importance, as it will be argued in the next section. As we mentioned above, the line shifts were displaced arbitrarily so that the strongest lines have zero shifts. We define the set of strong lines as those belonging to the plateau, where it is detectable. In other cases we assume that lines with $EW>150$~m\AA\ are ``strong'' in this context.

For the reference star, HIP~86400, the convective blueshifts are about $-150$~\ms for lines of $EW\simeq20$~m\AA. The blueshifts decrease (i.e., the line shift becomes less negative) as we approach $EW\simeq100$~m\AA. Beyond this $EW$ value, the shift is independent of $EW$.

The plateau is not clearly detected in HIP~16537, HIP~88601, HIP~64797, and HIP~37349. The latter is likely due to the low quality of the data compared to the rest of the sample stars. For the other stars in this list, this non-detection suggests that effects other than granulation shape the line profiles. Interestingly, except for HIP~26779, these are the same stars whose line bisectors have properties that differ from the theoretical expectation. As we suggested in Sect.~\ref{s:linebis}, the fact that these objects are also the most active among our sample stars may point to a correlation.

In setting the zero point of the convective blueshifts, we have assumed that the wavelengths of the cores of very strong \fei\ lines in K-dwarfs are not shifted. This is a reasonable assumption considering that these lines form in very high photospheric layers where the correlation between intensity and velocity fields is weak. Moreover, the assumption is justified by the fact that, in the solar case, where core wavelengths can be measured on an absolute scale, the plateau of convective blueshifts is found to be at zero velocity \citep{allende98:sun}.

Although there is large scatter for small $EW$ values, it appears that the line shifts become less dependent on $EW$ below $EW\simeq40$\,m\AA, suggesting another plateau (this is clearly visible in the case of HIP~96100). Interestingly, this result is somewhat consistent with our 3D model predictions, as it will be shown in \citetalias{kdwarfs-p2}.

\section{Impact on radial velocity determinations} \label{s:rv}

Although ultra high precision (close to the meter per second accuracy) can be achieved for relative measurements \citep[e.g.,][]{marcy98,queloz01,endl05}, absolute radial velocities are subject to a series of systematic errors that are very difficult to remove \citep{lindegren99,lindegren03}. The first of these is the gravitational redshift, which is on the order of 0.6~km~s$^{-1}$ in FGK stars. A secondary relativistic correction for the transversal Doppler effect is also required for high velocity stars, where it amounts to about 50~m~s$^{-1}$ \citep{lindegren99}. Both of these effects introduce a constant shift in the radial velocity zero point of the star. Equally or more important, however, is the effect of granulation.

If the cross-correlation method is used to determine the radial velocity of a star, then only when a very similar object is used as a radial velocity standard will the method be successful. This is due to the fact that the spectra of stars with different fundamental parameters show different line strengths, in addition to suffering from different granulation effects because of their different intensity and velocity field contrasts (Nordlund \& Dravins 1990), although the impact of the latter may be in general smaller. On the other hand, measuring the core wavelengths of a multitude of spectral lines and applying the Doppler formula to determine the stellar radial velocity introduces the error of the unknown convective blueshift to the measurements. However, as explained in Sect.~\ref{s:wavshifts}, the cores of very strong \fei\ lines in K-dwarfs have zero convective blueshift and, therefore, one can avoid the effects of granulation in radial velocity measurements by only using spectral lines whose convective line-shifts fall in the ``plateau'' region of the line-shift vs.~equivalent width relation, provided the plateau is reliably detected, as it is the case of some of our K-dwarfs: HIP~96100, HIP~86400, HIP~23311, HIP~88601, and HIP~26779 (cf.~Fig.~\ref{f:lshmk}).

The average values of the radial velocities we determined by cross-correlation of our cs21 spectra with each other and with their corresponding HET spectrum (see Sect.~\ref{s:observations} for details) are given in Table~\ref{t:rv}. While our relative velocities are very accurate (the relative velocities of cs21 pairs have a mean uncertainty of 12~m~s$^{-1}$ while that of a cs21-HET pair is about 70~m~s$^{-1}$), the HET reference spectrum has been only approximately corrected for the stellar radial velocity by using the Doppler shifts measured for a large number of strong spectral lines. The granulation effects are not yet removed from these data. Therefore, the radial velocities given in Table~\ref{t:rv} are only first order approximations, useful to check for large velocity variations, but should not be used as \textit{the} radial velocities of the stars.

\begin{table}
\centering
\begin{tabular}{lrrr} \hline\hline
Object & \multicolumn{1}{c}{$\mathrm{v}'_r$} & \multicolumn{1}{c}{$\sigma(\mathrm{v}'_r)$} & $n$ \\ 
       & \multicolumn{2}{c}{km s$^{-1}$} &  \\ \hline
HIP 96100  &  27.1	& 0.1 &  31 \\ 
HIP 86400  & 23.3 & 4.2 & 72 \\
HIP 23311  &  21.9	& 0.1 &  41 \\ 
HIP 16537  &  16.4	& 0.3 &  30 \\
HIP 64797  &   7.9	& 0.3 & 106 \\
HIP 26779  &   1.7	& 0.2 &  53 \\ 
HIP 88601  &  -5.4	& 0.2 &  31 \\
HIP 114622 &  -18.0	& 0.1 &  30 \\
HIP 37349  &  -18.0	& 0.2 &  35 \\ \hline
\end{tabular}
\caption{Approximate values of the radial velocity obtained for our sample stars using our data. The last column corresponds to the total number of observations. HIP~86400 is a confirmed single-lined spectroscopic binary; its orbital parameters are given in Table~\ref{t:rv86}. These values are not intended to represent the most accurate radial velocity determinations (those are given in Table~\ref{t:absrv}); they only serve our purpose of looking for obvious short-term radial velocity variations.}
\label{t:rv}
\end{table}

Table~\ref{t:rv} shows that almost all of our sample stars have nearly constant radial velocity, which makes them useful as radial velocity standards. HIP~88601, however, is a very long period visual binary star, while HIP~26779 is a suspected long period radial velocity variable (see details below). Only HIP~86400, our reference star, shows large, relatively short-term, radial velocity variations. Its orbital parameters are derived in Appendix~\ref{s:orbit86}.

We determined absolute radial velocities for the non-variable stars HIP~96100 and HIP~23311 using measurements of core wavelength shifts of their plateau lines. We also determined absolute radial velocities for HIP~26779, a suspected radial velocity variable, and HIP~88601, which is the primary star of a visual binary system that has an orbital period of 88.3 years and a velocity semi-amplitude of about 3.4~km~s$^{-1}$ \citep{heintz88}, which can be useful to improve the orbital solution of the systems or look for second order variations in the future. The procedure is described in detail below.

As explained in Sect.~\ref{s:observations}, one HET spectrum was used as a radial velocity template on which the cs21 pieces of the spectrum were put together. This HET spectrum was first shifted to an approximate laboratory wavelength frame by using a velocity obtained by comparing the observed core wavelengths of a large set of atomic lines with those measured in the laboratory. At that point, we were not interested in the absolute radial velocity of the star but only required to have the spectrum within about 1~km~s$^{-1}$ of its actual radial velocity for the purposes of unequivocally identifying strong spectral features. We will denote this first approximation to the radial velocity of the star with $\Delta\mathrm{v}_r^\mathrm{HET}$ (see Table~\ref{t:absrv}). The barycentric correction for the Earth-Moon system orbit around the Sun ($\Delta\mathrm{v}_r^\mathrm{bary}$) was then determined using the method outlined by \cite{mccarthy95}. The maximum error for this correction is expected to be about 5~m~s$^{-1}$ (M.~Endl; private communication).

In Fig.~\ref{f:lshmk}, the plateau lines have been put to zero velocity (on their average) by applying the $\Delta\mathrm{v}_r^\mathrm{plat}$ corrections given in Table~\ref{t:absrv}. Since, in principle, the plateau lines have zero convective blueshifts, this correction is precisely what is needed to convert $\Delta\mathrm{v}_r^\mathrm{HET}+\Delta\mathrm{v}_r^\mathrm{bary}$ into the absolute radial velocity of the star, but still uncorrected for relativistic effects ($\mathrm{v}_r^\mathrm{cl}$). Note that the standard deviation $\sigma$ from the mean estimate of $\Delta\mathrm{v}_r^\mathrm{plat}$ is between about 30 and 60~m~s$^{-1}$. Assuming that the line shifts of very strong lines in K-dwarfs are not affected by any other systematic error, the standard error $\sigma/\sqrt{n_L}$ would be an appropriate estimate of the uncertainty of this correction ($n_L$ being the number of strong lines used to determine the average value). Thus, after including the uncertainty of the barycentric correction, the error in our estimate of $\mathrm{v}_r^\mathrm{cl}$ is only about 10~m~s$^{-1}$.

To finally determine the absolute radial velocity of these stars, the relativistic gravitational redshift must be removed. This correction is given by $\Delta\mathrm{v}_r^\mathrm{rel}=-c\Delta\lambda/\lambda=[-636(M/R)+3]$~m~s$^{-1}$, if $M$ and $R$ are given in solar units. The +3 term corrects for the effects of the Earth's gravitational field \citep[e.g.,][]{dravins99}. We estimated the masses and radii of the four stars listed in Table~\ref{t:absrv} from the theoretical isochrones by \cite{bertelli94}, using the methods described in \cite{reddy03} and \cite{allende04:s4n}. Using the mass-to-radius ratios inferred from these calculations, we determined the $\Delta\mathrm{v}_r^\mathrm{rel}$ values, which allowed us to determine the absolute radial velocities, $\mathrm{v}_r$, given in the last column of the lower section of Table~\ref{t:absrv}.

\begin{table}
\centering
\begin{tabular}{ccrccc}  \hline\hline
HIP & JD-2454000 & $\Delta\mathrm{v}_r^\mathrm{HET}$ & $\Delta\mathrm{v}_r^\mathrm{bary}$ & $\Delta\mathrm{v}_r^\mathrm{plat}$ (std. error) \\ \hline
96100 & 231.936693 & 25.45 &  1.56 & $-0.07\pm0.04(0.01)$ \\
23311 & 406.891895 & 7.58 & 14.35 & $-0.01\pm0.03(0.01)$ \\
88601 & 205.922914 & $-29.54$ & 24.81 & $-0.03\pm0.06(0.01)$ \\
26779 & 350.969887 & $-23.79$ & 25.19 & $-0.15\pm0.06(0.01)$ \\ \hline

\\
\end{tabular}
\begin{tabular}{crccr} \hline\hline
HIP & \multicolumn{1}{c}{$\mathrm{v}_r^\mathrm{cl}$} & $\Delta\mathrm{v}_r^\mathrm{rel}$ & $\mathbf{v_r}$ \\ \hline
96100 & $26.93\pm0.01$ & $-0.68\pm0.06$ & $\mathbf{26.25\pm0.06}$ \\
23311 & $21.92\pm0.01$ & $-0.71\pm0.05$ & $\mathbf{21.21\pm0.05}$ \\
88601 & $-4.76\pm0.01$ & $-0.65\pm0.06$ & $\mathbf{-5.41\pm0.06}$ \\
26779 & $ 1.25\pm0.01$ & $-0.71\pm0.06$ & $\mathbf{\ \ 0.54\pm0.06}$ \\ \hline
\end{tabular}
\caption{Radial velocities of four of our sample stars. The several corrections are explained in detail in the text. The last column in the bottom table gives our best estimate of their absolute radial velocities. All velocities are given in km~s$^{-1}$. JD is the barycentric Julian date.}
\label{t:absrv}
\end{table}

From the calculations presented above, we conclude that only after the granulation effects have been properly taken into account, which implies reliably detecting the plateau in the line shift vs.~equivalent width relation and accurately measuring the line shifts of the spectral lines that define the plateau, the uncertainty in the absolute radial velocity of a star is dominated by the error in the gravitational redshift.

\cite{nidever02} have determined the absolute radial velocities of a large sample of late-type stars, including HIP~96100, HIP~23311, and HIP~26779. Their approach combines the classical method of cross-correlation with standard star templates with the very accurate relative radial velocity determination techniques used in spectroscopic exoplanet detection. Excluding systematic errors, they claim to achieve a precision of about 20--30~m~s$^{-1}$. For HIP~96100, HIP~23311, and HIP~26779, they derived 26.69, 21.55, and 1.21~km~s$^{-1}$, respectively. The differences with our values correspond to 0.44, 0.35, and 0.67~km~s$^{-1}$, with our velocities being smaller in all cases. \citeauthor{nidever02} estimated their systematic error due to be about +0.30~km~s$^{-1}$ in K-dwarfs, which, if considered, would improve the agreement between our results and theirs, with the possible exception of HIP~26779, for which the large difference that remains suggests long-period variability.

The high stability of the radial velocities of HIP~96100 and HIP~23311 make them good targets for radial velocity standardization. According to our calculations, the \textit{absolute} radial velocities of these K-dwarfs are known with an accuracy of about 60~m~s~$^{-1}$. Furthermore, for HIP~96100, \cite{gray92} have shown that the line bisectors are time independent, which suggests that the line shifts (and therefore the granulation correction to the absolute radial velocity) could also be constant, given that they are both observable manifestations of the same phenomenon.

\section{Summary and conclusions}

In this work, the signatures of granulation on the spectra of a small sample of K-dwarfs were detected using very high resolution ($R\simeq160,000-210,000$), high signal-to-noise ($S/N\gtrsim300$) observations with a spectral coverage from 5580 to 7800\,\AA. Not only were the data of very high quality, but also the reduction and post-reduction processing was carefully designed to preserve the relatively small effects of surface inhomogeneities. We were able to determine relative radial velocity shifts with a mean accuracy of 12~m~s$^{-1}$ for pairs of spectra of the same object, without requiring the spectrograph to be placed in a vacuum chamber or superimposing iodine features on the stellar spectra. The accurate determination of relative shifts was crucial for coadding multiple exposures of the same object to increase the signal-to-noise ratio of the observations while keeping the data useful for the detection and measurement of the relatively weak granulation signatures, which require very high $S/N$. However, we cannot guarantee that these values correspond to the real relative velocity variations of the stars because of shifts that could have been introduced by instrumental imperfections and/or environmental effects (e.g., variations of the air properties in the spectrograph room). In addition, we used a method of coaddition such that the interpolation distances in the resampling of frames before coadding was reduced by a factor of 2 compared to the standard method, thus reducing the impact of interpolation errors.

Using our carefully processed spectra, we measured the bisectors of a large number of \fei\ lines and averaged them out in groups of similar line-strength to minimize the impact of blends and noise. The observed mean bisectors show a characteristic C-shape, although for the strongest lines they have a shape that resembles only the lower part of the letter C (sometimes also referred to as a backslash shape: $\backslash$), which spans up to about 100~m~s$^{-1}$, even though the exact value depends on the particular K-dwarf star under consideration. We find that the stellar projected rotational velocity and activity influence the shapes of the line bisectors, reducing their magnitudes compared to inactive stars with low $V\sin i$ values. Although the effect of the projected rotational velocity is well understood, the details of the dependence on stellar activity, which we conclude it exists based on a correlation of level of chromospheric emission with line bisector span, remain open to investigation.

Core wavelength shifts were also determined for all our K-dwarf sample stars and they were found to show the typical properties of stellar granulation; for example, the weakest lines, which are formed in deep regions of the photosphere, where the correlation between temperature and velocity fields is stronger, show the largest blueshifts (around $-150$~m\,s$^{-1}$). Furthermore, the blueshifts decrease (i.e., the core wavelength shift, by definition positive for redshift, increases) for stronger lines and, for some of our sample stars, they appear to reach a constant value for \fei\ lines stronger than $EW\simeq100$~m\AA. We interpret this plateau as line formation taking place in high photospheric layers, where the correlation between intensity and velocity fields is very weak or null and therefore the convective line shifts are expected to be nearly zero. This behavior had been identified before only in the solar spectrum. The detection of a plateau allows us to determine the zero point of the convective blueshifts, which is necessary to remove the large uncertainties due to granulation in the determination of absolute stellar radial velocities. In some of our K-dwarfs, the plateau could not be reliably detected, most likely due to extrinsic effects such as observational noise, scatter due to large rotational velocities, or stellar activity.

Thanks to our quantitative understanding of the effects of granulation on the core wavelengths of \fei\ lines in the spectra of K-dwarfs, we have been able to determine the absolute radial velocities of two stars that have non-variable radial velocities and a very well defined plateau in the convective blueshift vs.~line strength relation (HIP~96100 and HIP~23311). These extremely accurate measurements make these stars, together with the Sun and asteroids with known orbits, the best radial velocity standards available. In addition, we have measured radial velocities with similar precision for two stars that show long-term variability: HIP~88601, which is the primary star of a visual binary system, and HIP~26779, a suspected radial velocity variable.

The apparent non-detection of the plateau in other sample stars does not imply that granulation on their surfaces behaves differently, given that they suffer from other phenomena that preclude detection. Thus, in principle, it is possible to remove the errors due to granulation on measurements of absolute radial velocities of many K-type dwarf stars, which could be extremely useful for future massive spectroscopic surveys, where millions of K-dwarfs can be used to establish an accurate zero point for the stellar radial velocities.

\begin{acknowledgements}
This work was supported in part by the Robert A. Welch Foundation of Houston, Texas. We thank Mike Endl for providing the barycentric corrections for the Earth-Moon system used in our determination of absolute radial velocities and the anonymous referee for his/her constructive criticism. Some of the data used in this work were obtained at the Hobby-Eberly Telescope (HET), which is a joint project of the University of Texas at Austin, the Pennsylvania State University, Stanford University, Ludwig-Maximilians-Universit\"at M\"unchen, and Georg-August-Universit\"at G\"ottingen. The HET is named in honor of its principal benefactors, William P. Hobby and Robert E. Eberly. We are indebted to D. Doss as well as the HET and McDonald Observatory staffs for their support during our observations.
\end{acknowledgements}

\bibliographystyle{aa}
\bibliography{ir_refs}

\begin{appendix}

\section{Projected rotational velocities} \label{s:vsini}

Most of our sample stars have one or more entries in The Catalog of Stellar Projected Rotational Velocities \citep{glebocki00}, which is a compilation of $V\sin i$ values available in the literature and, therefore, determined using a number of different techniques such as Fourier transform of line-profile (FTPL), cross-correlation (C-C), calibrated line-width at half maximum (FWHM), and convolution with calculated rotational broadening (Conv). Table~\ref{t:vsini} lists the values given in this catalog for our sample of K-dwarfs (the original references are also provided). Entries giving only upper or lower limits as well as very old and uncertain values have been excluded. Note that \cite{fekel97} calibrated his method using measurements by D.~F.~Gray and his results are therefore not independent from, for example, those given in \cite{gray84:rotation}. In addition, \citeauthor{fekel97}'s assumption of a constant macroturbulent velocity for K-dwarfs makes those results less reliable. The $V\sin i$ values determined by \cite{valenti05} are also given in Table~\ref{t:vsini}.

By convolving line profiles predicted by a 3D K-dwarf model atmosphere with projected rotational velocity profiles, as it will be described in detail in Paper II, we determined a more accurate $V\sin i$ value for HIP~86400. The result is given in Table~\ref{t:vsini}. Proper disk integrations were compared against the convolution method and the line profiles computed with both approaches were found to agree remarkably well \citep[][pp.~88--90]{ramirez08:thesis}. We did not perform similar calculations for the other sample stars because the line profiles are very sensitive to the stellar parameters, in particular effective temperature, and only HIP~86400 has parameters identical (within observational errors) to those of the 3D model.

\begin{table}

\centering

\begin{tabular}{lccl} \hline\hline
Object         & $V\sin i$ & Technique & Reference \\ 
               & (km~s$^{-1}$) &       &         \\ \hline
HIP 96100      &  $0.6\pm0.8$ & FWHM & \cite{fekel97} \\
	       &  $0.8\pm0.4$ & FTLP & \cite{gray84:rotation} \\
	       &  $1.4\pm0.5$ & Conv & \cite{valenti05} \\ \hline
HIP 26779      &  $4.0\pm0.6$ & C-C  & \cite{gaidos00} \\
	       &  $3.3\pm0.5$ & Conv & \cite{valenti05} \\ \hline
HIP 16537      &  $1.5\pm2.0$ & C-C  & \cite{benz84} \\
	       &  $2.0\pm0.8$ & FWHM & \cite{fekel97} \\
	       &  $2.2\pm0.4$ & FTLP & \cite{gray84:rotation} \\
	       &  $1.7\pm0.3$ & Conv & \cite{saar97:south} \\
	       &  $1.4\pm1.0$ & Conv & \cite{saar97:south} \\
	       &  $1.0\pm0.5$ & FTLP & \cite{smith83} \\
	       &  $2.5\pm1.3$ & C-C  & \cite{tokovinin92} \\
     	       &  $2.4\pm0.5$ & Conv & \cite{valenti05} \\ \hline
HIP 88601      &  $3.1\pm0.8$ & FWHM & \cite{fekel97} \\
               &  $3.3\pm1.0$ & FTLP & \cite{smith78} \\
               &  $1.6\pm0.4$ & FTLP & \cite{gray84:rotation} \\ \hline
HIP 64797      &  $3.9\pm0.8$ & FWHM & \cite{fekel97} \\
	       &  $2.8\pm0.6$ & Conv & \cite{hale94} \\ \hline
HIP 37349      &  $2.5\pm0.5$ & Conv & \cite{valenti05} \\ \hline
HIP 86400      &  $2.6\pm2.0$ & C-C  & \cite{benz84} \\ 
       	       &  $1.6\pm0.2$ & Conv* & TW \\ \hline
HIP 114622     &  $2.1\pm0.8$ & FWHM & \cite{fekel97} \\
	       &  $1.8\pm0.5$ & Conv & \cite{valenti05} \\ \hline
HIP 23311      &  $0.8\pm0.8$ & Conv & \cite{saar97:south} \\
               &  $1.4\pm1.8$ & C-C  & \cite{tokovinin92} \\
	       &  $1.7\pm0.5$ & Conv & \cite{valenti05} \\ \hline
\end{tabular}	
	
\caption{$V\sin i$ values reported in the literature for our K-dwarf sample stars. The value determined for HIP~86400 in Paper~II of this work (TW) is also given.}

\label{t:vsini}

\end{table}

\section{Chromospheric activity} \label{s:activity}

\begin{figure}
\includegraphics[width=9cm,bb=45 365 480 870]{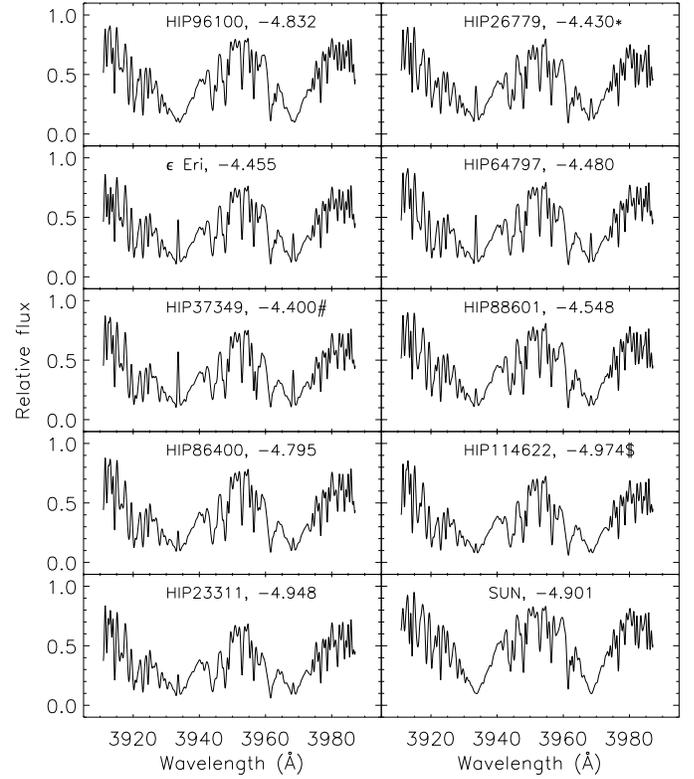}
\caption{Medium resolution spectra of our K-dwarfs and the Sun in the region around the Ca~\textsc{ii} H and K lines. The activity index $R_\mathrm{HK}'$ is shown in each panel, as given by \cite{saar99}, except for the cases marked with the following symbols: [*] = \cite{rocha98}, [\#] = \cite{king03}, [\$] = \cite{saar97:cool}.}
\label{f:activ}
\end{figure}

Many K-type dwarf stars reveal high levels of chromospheric activity in the form of emission in the cores of very strong lines \citep[e.g.,][]{noyes84}. This may have an important impact on the shapes of spectral lines in general. In Fig.~\ref{f:activ}, we show the spectral region around the Ca~\textsc{ii} H and K lines for all our sample stars except Procyon. The data were acquired from the spectroscopic survey of \cite{allende04:s4n} but we have reduced the resolution to about $R=8,000$ for clarity in the illustration. The chromospheric activity index $R'_\mathrm{HK}$ for each star is also given in Fig.~\ref{f:activ}. This information is relevant to the discussion presented in Sect.~\ref{s:measurement}.

\section{Orbital solution for the single-lined spectroscopic binary HIP~86400} \label{s:orbit86}

\begin{figure}
\includegraphics[width=9cm,bb=70 370 450 830]{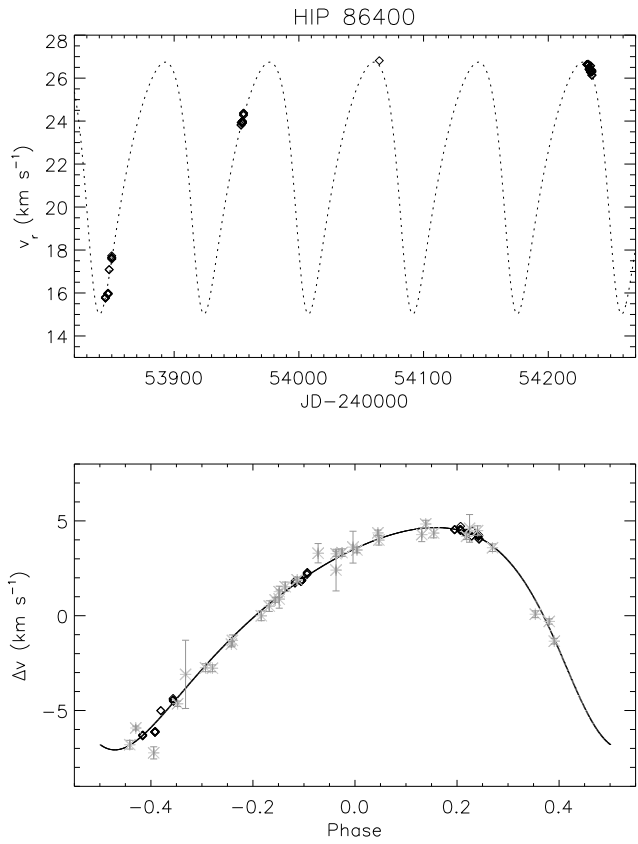}
\caption{Upper panel: radial velocities of HIP~86400 obtained from our observations. The dotted line is a model fit to the data. Lower panel: fit to our radial velocity data (black diamonds) and those of Tokovinin (1991; gray asterisks with error bars). The orbital parameters are given in Table~\ref{t:rv86}.}
\label{f:rv86}
\end{figure}

Combining our radial velocity data with those of \cite{tokovinin91}, we were able to fit a spectroscopic orbit for the primary star of the HIP~86400 system using the \textsc{GaussFit} package \citep{jefferys88}. The resulting orbital parameters are given in Table~\ref{t:rv86} and the model fit to the data is shown in Fig.~\ref{f:rv86}.

The mass function of the system is $0.00153\pm0.00008M_\odot$, which implies a mass of about $0.42M_\odot$ for the secondary, assuming a mass of $0.85M_\odot$ for the primary (determined from isochrones) and $\sin i=0.32$ (as measured astrometrically by \citealt{jancart05}). This corresponds to an M-dwarf star with an effective temperature around 3500\,K \citep{baraffe96,basri00} whose contribution to the flux is relatively small in the visible and near infrared (a secondary spectrum has not been detected yet). Using the spectral energy distributions calculated from Kurucz model atmospheres, we find that the surface flux from the M-dwarf in the HIP~86400 system is about 8\% of the flux emitted by the K-dwarf in the spectral region of interest for this work. Considering that the flux received on Earth is proportional to the angular diameter of the star squared and the radii of $0.4M_\odot$ dwarfs are about $0.4R_\odot$ \citep{beatty07}, the contribution to the flux from the M-dwarf to our observed spectrum of HIP~86400 is about $(0.4/0.8)^2\times8$\%$=2$\,\%, given that the radius of the primary, the K-dwarf, is about $0.8R_\odot$.

\begin{table}
\begin{tabular}{lcl} \hline\hline
Parameter & Value & Units \\ \hline
Orbital period ($P$) & $83.713\pm0.005$ & days \\
Time of periastron passage ($T$)      & $2447723.54\pm0.47$ & JD \\
Eccentricity ($e$) & $0.288\pm0.012$ & \\
Longitude of periastron ($\omega$) & $135.99\pm1.97$ & degrees \\
Semi-amplitude of RV curve ($K_1$) & $5.86\pm0.10$ & km~s$^{-1}$ \\
Velocity of the center of mass ($\gamma_a$) & $22.095\pm0.077$ & km~s$^{-1}$ \\
Velocity of the center of mass ($\gamma_b$) & $21.194\pm0.070$ & km~s$^{-1}$ \\
Mass function ($f_m$) & $0.00153\pm0.00008$ & $M_\odot$ \\ \hline
\end{tabular}
\caption{Orbital parameters obtained by fitting the radial velocity data of HIP~86400. Two values for $\gamma$ are given; the first one, $\gamma_a$, corresponds to our observations, while $\gamma_b$ was obtained for Tokovinin (1991) data. A systematic difference of about 0.9~km~s$^{-1}$, with our velocities being larger, is present between the two data sets. The rest of the parameters were derived using the combined data set. The radial velocity curve of HIP~86400 is shown in Fig.~\ref{f:rv86}.}
\label{t:rv86}
\end{table}

\end{appendix}

\end{document}